\documentclass[onecolumn]{elsart}    % Enable this line and disable the
\usepackage{amssymb,amsmath}
\usepackage{latexsym}
\usepackage{amsfonts}
\usepackage{graphicx}%
\usepackage{epsfig}

\renewcommand{\o}{\omega}
\def\beq{\begin{equation}}
\def\eeq{\end{equation}}
\def\beqa{\begin{eqnarray}}
\def\eeqa{\end{eqnarray}}

\def\o{\omega}

\begin{document}

\begin{frontmatter}

\title
{A Two Phase Harmonic Model for Left Ventricular Function}

\author[d1]{Shay Dubi \thanksref{PhD} }\ead{shay@naiot.com}
 \thanks[PhD]{This work was performed in partial fulfilment of the 
requirements for a Ph.D. degree of S. Dubi, Sackler Faculty of Medicine, Tel Aviv University, Israel},    % Add the
\author[d2]{Chen Dubi}\ead{chend@sce.ac.il},               % e-mail address
\author[d3]{Yonatan Dubi}\ead{dubij@bgu.ac.il}  % (ead) as shown

\address[d1]{Sackler Faculty of Medicine, Tel Aviv University, Israel}  % Please supply
\address[d2]{Unit of Mathematics, Negev Academic College, Basel 9 Beer Sheva, Israel}             % full addresses
\address[d3]{Physics Department, Ben-Gurion University, Beer Sheva 84105, Israel}        % here.

%\author{Shay Dubi$^1$, Chen Dubi$^{2}$ and Yonatan Dubi$^{3}$}
%\author{Shay Dubi\footnote{Department of Hartabuna, Tel-Aviv University, Tel-Aviv ***Add address***}
%, Chen Dubi\footnote{Unit of Mathematics, Negev Academic College, Basel 9 Beer Sheva, Israel}  and Yonatan Dubi\footnote{Physics Department, Ben-Gurion University, Beer Sheva 84105, Israel}}
%\author{Yonatan Dubi}
%\affiliation{
%$^{1}$ Department of Hartabuna, Tel-Aviv University, Tel-Aviv ***Add adress***\\
%$^{2}$ Unit of Mathematics, Negev Academic College, Basel 9 Beer Sheva, Israel\\
%$^{3}$ Physics Department, Ben-Gurion University, Beer Sheva 84105, Israel
%}

                             % keyword list or with the

\date{}
\begin{abstract}
A minimal model for mechanical motion of the left ventricle is proposed.
 The model assumes the left ventricle to be a harmonic
oscillator with two distinct phases, simulating the systolic
 and diastolic phases, at which both the amplitude and the elastic
constant of the oscillator are different. Taking into account the
 pressure within the left ventricle, the model shows qualitative
agreement with functional parameters of the left ventricle. The model
 allows for a natural explanation of heart failure with
preserved systolic left ventricular function, also termed diastolic heart failure.
 Specifically, the rise in left ventricular filling
pressures following increased left-ventricular wall stiffness is attributed
 to a mechanism aimed at preserving heart rate and
cardiac output.
\end{abstract}

\end{frontmatter}

%\begin{quote}
%\end{quote}
%\tableofcontents
%\vskip 1truecm \baselineskip=14pt
%\parindent 0cm
%\newtheorem{Pa}{Paper}[section]
%\newtheorem{Tm}[Pa]{{\bf Theorem}}
%\newtheorem{La}[Pa]{{\bf Lemma}}
%\newtheorem{Cy}[Pa]{{\bf Corollary}}
%\newtheorem{Rk}[Pa]{{\bf Remark}}
%\newtheorem{Pn}[Pa]{{\bf Proposition}}
%\newtheorem{Pb}[Pa]{{\bf Problem}}
%\newtheorem{Dn}[Pa]{{\bf Definition}}
%\newtheorem{Ex}[Pa]{{\bf Example}}
%\renewcommand{\theequation}{\thesection.\arabic{equation}}
%\setcounter{equation}{0}
\begin{center}

\end{center}
\section{Motivation}
\subsection{General}
The left ventricle (LV) is a highly complex mechanical system. Left 
ventricular function is determined by both internal and external factors 
such as myocardial contractility, myocardial stiffness, preload, afterload, rhythm and pace,
size, shape etc \cite{Starling}.  % Lionel H. Opie. Mechanisms of cardiac contraction and relaxation. In: 
%Zipes D.P., Libby P., Bonow R.O., Braunwald E., editors. Braunwld's heart disease, a textbook of 
%cardiovascular medicine, 7th edition. Philadelphia, Pennsylvania, Elsevier Saunders; 2005. p. 473-487.   ****. 
All these factors combine to regulate left ventricular
motion, which may be characterized by three basic distinguishing attributes:
(1) the motion is periodic, including a diastolic expansion phase and a systolic contraction phase, 
(2) the period is divided unequally between the diastolic and the systolic phases,
the former being generally twice as long as the latter \cite{Tortora} %Tortora G.J., Derrickson B. Principles of anatomy and physiology 11th edition.
%Hoboken, NJ, John Wiley & Sons;2006. p. 716. ****, 
and, 
(3) the forces generated by the left ventricle are in accordance with Frank-Starling 
law of the heart \cite{Starling}, ], which states that the larger the LV volume, the greater the energy 
of it's contraction
 and hence diastolic stretch actually increases LV contractility. 

Motivated by the fact that LV dysfunction and heart failure are associated with alterations
 in the mechanical motion of the left ventricle \cite{HFa,HFb,HFc}
%(Colucci W.S. Braunwald E. Pathophysiology of heart failure. In: 
%Zipes D.P., Libby P., Bonow R.O., Braunwald E., editors. Braunwld's heart disease, a textbook of 
%cardiovascular medicine, 7th edition. Philadelphia, Pennsylvania, Elsevier Saunders; 2005. p. 509-526.) (Zile MR, Brutsaert DL. New concepts in diastolic dysfunction and diastolic heart failure: 
%Part I: Diagnosis, Prognosis, and Measurements of Diastolic Function. Circulation. 2002;105:1387–1393.
%Zile MR, Brutsaert DL. New concepts in diastolic dysfunction and diastolic heart failure: 
%Part II: causal mechanisms and treatment. Circulation 2002; 105(12):1503-1508.)  ***
, simulating this motion has become a major challenge. The use of both theoretical tools (such as finite element 
geometrical construction\cite{hunter}, fluid-dynamics based computation \cite{fluid1,fluid2}, Immersed Boundary method
 \cite{immersed1} etc.) and experimental
tools (such surgically implanted track markers \cite{exp1,exp2}, magnetic resonance tagging \cite{NMR,kindberg}
 etc.) have been utilized for
such simulation. As a common factor, many of the theories attempt to include 
an abundance of factors influencing the left ventricle, to achieve
quantitative resemblance between theory and experimental data. While some success has been achieved 
in describing normal LV function, the models are far from adequate in describing LV dysfunction 
or the effect of cardiac assist devices, and further progress is needed before such models
 may be utilized for clinical applications \cite{clinical}.

\subsection{Diastolic heart failure} \label{introDHF}
Heart failure (HF) is a clinical syndrome that can result from structural or functional cardiac disorders, impairing the
ability of the ventricle to fill with or eject blood \cite{Hunt}. The clinical manifestations of HF are dyspnea and fatigue,
which may limit exercise tolerance, and fluid retention, which may lead to pulmonary and peripheral oedema.
Diastolic heart failure (DHF) is the clinical syndrome of heart failure associated with preserved LV 
ejection fraction (an index of LV systolic function and contractility, defined as the ratio of LV stroke volume
 to end diastolic volume). DHF accounts for approximately 40\% of heart failure cases \cite{DHF1,DHF2} and carries 
a significant morbidity, comparable to that of systolic heart failure \cite{DHF3,DHF4}. 
Diastolic abnormalities of the LV may be related to increased 
myocardial stiffness and impaired relaxation \cite{DHF7,DHF7a,DHF7b}, leading to an essentially {\sl mechanical} dysfunction, 
resulting in the inability of the left ventricle to fill with blood at low filling pressures
during diastole \cite{DHF5,DHF6}.

 The mechanisms underlying increased myocardial 
stiffness can be divided into factors that are intrinsic or extrinsic to the myocardium \cite{DHF8}. Myocardial 
factors include cellular factors, such as impaired calcium homeostasis and changes in Titin sarcomeric proteins, which 
act as  viscoelastic springs that provide a recoiling force during diastole \cite{DHF9}, and extracellular factors, 
including changes in extracellular matrix morphology (i.e. fibrosis) leading to increased myocardial stiffness \cite
{DHF10}. Relaxation is an energy-dependent process, and abnormalities in cellular energy supply and utilization can 
lead to impaired relaxation \cite{DHF11}. Both abnormalities may result in a physiological state in which diastolic LV 
pressures are elevated, leading to elevated left atrial and pulmonary venous pressures, exercise intolerance \cite
{DHF12,DHF12a} and acute pulmonary oedema \cite{DHF13,DHF13a}.

Despite its prevalence and significant morbidity there are currently very few models that can 
simulate DHF as an integral part of the mechanical motion of the LV \cite{dysfunction}. Such simulation would 
be beneficial for promoting our understanding of the underlying pathophysiology of DHF and 
simulating treatment modalities, 
and as such provided much of the motivation for developing a minimal model for LV motion. 

\subsection{Scope}
In this paper we develop a qualitative minimal approach to
describe the motion of the LV. Generalizing the simple linear
harmonic oscillator, we qualitatively mimic the full ventricular cycle, with all three basic distinguishing 
attributes described above. Further more, our model naturally describes the mechanics
of DHF, and thus provides a baby-step towards a more rigorous model for LV function in DHF.  

The rest of this paper is organized as follows. In Sec.~\ref{model} we describe and justify our
model. In Sec.~\ref{DHF} we discuss the relation between our model and DHF, and Sec.~\ref{summary} is devoted to a summary.

\section{The model}\label{model}
\subsection{Preliminaries}
As a mathematical introduction, in this section we discuss the
simple harmonic oscillator, describe the general solution and show
its relation to the properties of the LV. A simple harmonic
oscillator (e.g. a mass attached to a spring, an elastic band
etc.) is described by Hooke's law, that is the larger the
displacement from equilibrium the stronger the restoring force
acting on the body. In the simplest case, the restoring force is
proportional to the displacement, and thus Newton's law is of the
form \beq r''(t)=-K r(t) \label{newton2}~~, \eeq where $K$ is the
proportionality (elastic) constant between the force and the displacement,
its physical meaning is the amount of force required to 
stretch the oscillator by one unit length (per unit mass).

 The solution of Eq.~\eqref{newton2} depends on the sign of $K$.
If $K$ is negative, the general solution is \beq r(t) = r_{+}
\exp(\sqrt{-K} t)+r_{-} \exp(-\sqrt{-K} t) ~~,\label{sol1}\eeq where
$r_{+}$ and $r_{-}$ are constants determined by the initial
conditions of the oscillator. Notice that this solution is
"unstable", that is $r(t)$ will increase indefinitely with the
time $t$. Thus, this solution is inappropriate for describing an
oscillating body.

 For a positive $K$,
the solution is periodic, \beq r(t)=r_0 \cos( \sqrt{K} t + \phi)
~~, \label{sol2}\eeq where the radial amplitude $r_0$ and the phase $\phi$ are constants, which
are determined by the initial conditions of the oscillator. The
frequency of oscillations is given by $\o = \sqrt{K}$, and the
period (i.e. the duration of a single cycle) is $t_s= 2 \pi / \o$.
 We notice that if $r(t)$ stands for the LV radius, than this solution already possesses two of the 
LV characteristics described in the introduction. Obviously, the motion is periodic, and the 
force (given by the right hand side of Eq.~\ref{newton2}) is in agreement with Starlings law.  

However, the third characteristic of the LV - the unequal division of
the cardiac cycle - is unfulfilled by the simple harmonic oscillator.
Following, we introduce a generalization to the simple oscillator,
based on a physiologic mechanism, which inherently incorporates this feature.

\subsection{Basic model}

In this section we describe our generalization of the simple
harmonic oscillator. We start by examining the physiologic origin
of systolic and diastolic LV function.

The transition from diastole to systole stems
from excitation-contraction coupling. The trigger for systole is the electrical excitation of 
cardiac muscle cells (action potential), which raises the free intra-cellular $Ca^{2+}$ concentration,
 thus activating cellular contraction.
Contraction leads to elevation of ventricular wall tension, elevation of intra-ventricular pressure
and eventually ejection of blood from the ventricle, when a positive pressure gradient is achieved. 
For diastolic relaxation to occur, intra-cellular $Ca^{2+}$ concentration must be actively reduced, 
leading to myocardial relaxation \cite{cycle}. 
% Source:  Cardiac excitation–contraction coupling. Donald M. Bers, NATURE | VOL 415 | 10 JANUARY 2002

In essence, this can be simplified to a model in which systole begins when the LV wall
is abruptly stiffened (isovolumic contraction) leading to LV pressure elevation and systolic contraction, after-which 
the ventricle is suddenly relaxed (isovolumic relaxation), followed by a pressure decline and diastolic filling.  
Even a model based on this simplification would require consideration of additional factors such as, e.g.
environmental effects, blood flow (rheological) effects, spatial form of the LV, energy considerations etc.). This type of 
analysis is probably beyond mathematical rigour. 

We thus suggest a phenomenological model for LV function.
We describe the LV as a cylindrical elastic membrane. The model
thus consists of a harmonic oscillator, for which both the elastic
constant $K$ (and thus the frequency) and the radial amplitude $r_0$ differ
between the systolic and the diastolic phase. Since the systolic
(diastolic) phase are defined by a decrease (increase) of the
LV radius $r(t)$, {\sl both the elastic constant and the amplitude 
will depend on the sign of the derivative}, $r'(t)$ . This will
mimic the sudden stiffening and relaxation of the cardiac muscle due to changes
 in intra-cellular $Ca^{2+}$ concentration.
The force induced by the intra-ventricular pressure, $F$, is given (in the
cylindrical approximation) by $F(t)=2 \pi r(t) P(t)$, where $P(t)$
is the internal pressure (divided by the LV wall mass and height). 
Assuming that in the systolic and diastolic phases
the pressures $P_s$ and $P_d$ are approximately constant, we obtain the following equation for the
radius of the LV,
\begin{equation}\label{Shahar1}
\begin{cases}
r''(t)+K_{s}(r-r_{0,s})+2\pi r P_{s}=0, \quad
r'(t)<0 \quad \\ \hskip 5.5truecm \text{(systolic phase)}\\
r''(t)+K_{d}(r-r_{0,d})+2\pi r P_{d}=0, \quad
r'(t)>0 \quad \\ \hskip 5.5truecm \text{(diastolic phase)}
\end{cases}
\end{equation}
where $K_{d}$ and $K_{s}$ are the effective elastic constants at the diastole and systole respectively, and 
$r_{0,d}$ and $r_{0,s}$ are the amplitudes at the diastole and systole, respectively. Eq.~\ref{Shahar1} is   
subjected to the initial conditions
\begin{equation}\label{salvation}
r(0)=r_{max}; \quad r'(0)=0
\end{equation}
where $r_{max}$ is the maximal radius of the LV, 
and to the following conditions:
\begin{enumerate}
\item $r(t)$ is a continuous function.
\item The motion is periodic; that is, if $T$ is the period
of the function, then
\[
r(0)=r(T)=r_{max}.
\]
\end{enumerate}
Let us start by examining each branch of the solution separately, starting 
with Eq.~\eqref{Shahar1} for the the systolic phase. Demanding that
the solution is stable yields the condition $K_{s}>2\pi P_{s}$. Denoting
$\frac{K_{s}r_{0,s}}{K_{s}-2\pi P_{s}}=r_{s}$ and taking in
account the initial conditions (Eq.~(\ref{salvation})), we have the
 solution
\begin{equation}\label{R-Systola}
r(t)=(r_{max}-r_{0})\cos \left(\sqrt{K_{s}-2\pi P_{s}} t\right)+r_{0}.
\end{equation}
The motion is
governed by Eq.~\eqref{R-Systola} as long as the first derivative of the
solution is negative. The derivative changes sign (from negative to
positive) at $t_s=\frac{\pi}{\sqrt{K_{s}-2\pi P_{s}}}$, for which $r(t_s)=(r_{max}-r_{0})\cos\pi +r_{0}=2r_{0}-r_{max}$.
Thus, $r_{min}=2r_{0}-r_{max}$ (where $r_{min}$ is the minimal LV radius) is the initial condition for the motion in the diastolic phase, i.e. the solution
for the diastolic part of Eq.~(\ref{Shahar1}) is subjected to the initial conditions,
\begin{equation}\label{Vick}
r(t_s)=r_{min};\quad r'(t_s)=0.
\end{equation}
Denoting $r_{d}=\frac{K_{2}r_{0,d}}{K_{2}-2\pi P_{d}}$, the
solution for the diastolic stage is given by
\begin{equation}\label{R-Diastola}
r(t)=(r_{min}-r_{d})\cos \left( \sqrt{K_{d}-2\pi P_{d}} (t-t_s) \right) +r_{d}.
\end{equation}
As in the systolic phase, this branch of the solution is valid when $t$
is in the interval $t_s \leq t \leq t_s+\frac{\pi}{\sqrt{K_{d}-2\pi P_{d}}}
$. Denoting $t_d=\pi / \sqrt{K_{d}-2\pi P_{d}} $, $\o_{s}=\sqrt{K_{s}-2\pi P_{s}}$ and
$\o_{d}=\sqrt{K_{d}-2\pi P_{d}}$
we obtain the solution for a single cycle,
\begin{equation}\label{FullCycle}
r(t)=\begin{cases}
                (r_{max}-r_{s})\cos (\o_{s} t) +r_{s}
                         ; \quad 0 \leq t \leq t_s \\
                (r_{min}-r_{d})\cos( \o_{d}(t-t_s))+r_{d}
                         ;  \quad t_s \leq t \leq t_s+t_d \\
     \end{cases}
\end{equation}
From continuity at $t=t_s$ and periodicity (i.e. $r(t_s+t_d)=r(0)$) we
obtain the relation \beq\label{r0r1}
r_{s}=r_{d}=\frac{r_{max}+r_{min}}{2}. \eeq

We thus obtain a full
solution for the motion of the LV, characterized by the four
parameters $t_s,~t_d,~r_{max}$ and $r_{min}$. The ratio between $t_d$ (the duration of diastole) and $t_s$ 
(duration of systole) 
is $t_d /t_s \approx 2$. The parameters $r_{max}$ and $r_{min}$
may vary and are determined by both the elastic constants and the
pressures in the different phases.

\begin{figure}[h!]
\centering
\includegraphics[width=8truecm]{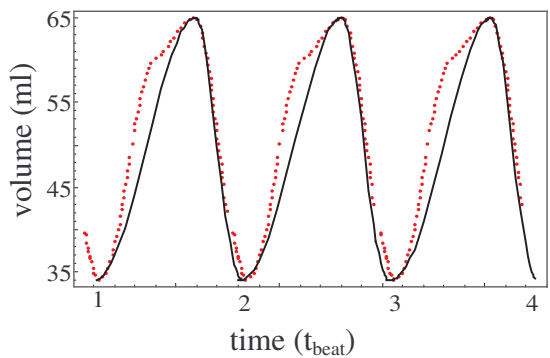}
%\parbox{12truecm}{
\centering
\caption{\footnotesize Dashed line: LV volume, data taken from a healthy human subject using non-invasive magnetic
 resonance imaging technique, adopted with permission from Ref.~\cite{kindberg}.  
Solid line : LV volume, $\propto \pi r(t)^2$, as extracted from the model,
 as a function of time (in units of a single cycle period, $t_{\mathrm{beat}}$).
 The fit between the model and data yields a ratio between the diastolic and systolic times, $t_d /t_s =2.152 $.}
%}
\label{r_t} \end{figure}

In Fig.~\ref{r_t} we plot the LV volume for several cycles,
  actual LV data taken from a healthy human subject using non-invasive magnetic
 resonance imaging technique, adopted from K. Kindberg \cite{kindberg} (dashed line) and a fit with
 Eq.~(\ref{FullCycle}) (solid line).
 As seen, the model yields fair agreement with the data, and results in the ratio
 between the diastolic and systolic times, $t_d /t_s =2.152 $. 

%\subsection{Estimation of parameters}
%In this section we evaluate $K_{s}$ and $K_{d}$ for a real heart.
%In real life measurement, a typical value of $P_{s}$ is $120$ mmHG , and
%a typical value of $P_{d}$ is $10$ mmHG. Let us denote by $N$ the number of beats per minute (BPM)
%of a given heart
%(standard healthy values for $N$ are $60 \leq N \leq 90$). In each
%heart beat the diastole is approximately twice as long as the
%systole, and hence the duration of a single systole is
%$t_s \sim 2 / 3N $ where as the duration of a single diastole is about
%$t_d \sim 1 / 3N $. Thus, by using Eq.~\eqref{FullCycle}
%$K_{s}$ and $K_{d}$ may be extracted. For instance, for a heart
%rate of $N=60$ BPM and pressures $P_{s}=110, P_{d}=10$, we have
%the numerical values
%\begin{equation}
%K_{s} \approx 779.977; \quad K_{d}=85.0385
%\end{equation}%

%******* What are these numbers ? Units ? $m \times$mmHG $\rightarrow 1/$sec ?
%****** Do they fit some real data ?
%****** measurable ? in what units ? has been measured ?

\section{Diastolic heart failure}\label{DHF}

As described in Sec.~\ref{introDHF}, diastolic abnormalities in DHF patients include increased LV stiffness and 
elevated LV filling pressures, leading to the symptoms of heart failure.  
In this section we intend to answer two questions:
\begin{enumerate}
\item Can we simulate the increased LV stiffness within our model?
\item Can this model predict and explain the increase of pressure?
\end{enumerate}
For the first question, the answer seems fairly simple: In the
suggested model, LV stiffness is represented by the
coefficients $K_{s}$ and $K_d$. Thus, in terms of the suggested model, the disease
can be simulated by increasing $K_{d}$ by an amount $\Delta K$.

In order to answer the second question we note that in DHF
 patients there seems to be no significant change in the heart rate, and thus
 no real change in the duration of the diastole (or systole) \cite{DHF13a,Kitzman}.
 Since the duration of the phases is defined by the frequencies $\o_{s}$ and
$\o_{d}$,  from their definition we have that $\o_{d}$ remains constant only if the term $K_{d}-2\pi P_{d}$
remains constant. Thus, if $K_{d}$ is increased by
a portion $\Delta K$, then $P_{d}$ must also increase by $\frac{\Delta K}{2\pi}$.
Hence, this model indeed predicts the rise in pressure. In other words, once the diastolic 
stiffness of the LV wall increases, a
compensation is achieved by increasing the pressure, in order to
prevent a significant change in heart rate.

%Following the above analysis, one would expect that once
%the stiffness of the heart is increased, it would affect both
%the systole and the diastole. If this is the case, why are the
%symptoms only visible only at the diastole? The answer is that
%indeed, the systolic pressure should change, and in the same
%amount as the diastolic pressure. However, since the initial
%systolic pressure is much higher, the relative change is minor.
%For example, in the acute case at which the diastolic pressure is
%doubled (say, from $10$ mmHG to $20$ mmHG), the relative change in
%systolic pressure is less than $10 \%$. ****** Look for data in literature, 
%in the change of pressures at DHF, and see if there is a detectable change in Ps***.

%Following this, we hypothesize that DHF,
%as a mechanical flaw, affects the heart in general- both the
%systole and the diastole. Yet, since the affect is barely noticeable
%in the systole, the disease is only attributed to the diastole.

\section{Summary}\label{summary}

In summary, a naive harmonic model for the mechanical motion
of the left ventricle was developed. The model consists of a generalized harmonic oscillator, 
for which both the radial amplitude and the elastic constant depend on the direction of the motion,
 yielding both the systolic and diastolic phases of the cardiac cycle, and allows for a
 qualitative simulation of left ventricular motion. 

Despite its simplicity, the model yields a natural explanation for
 diastolic heart failure, in which the left ventricular wall is stiffened and diastolic
 filling pressures are increased. Our model suggests that the rise in diastolic filling
 pressures is required in order to maintain ventricular volume when the heart rate is not
 significantly changed, as is the clinically observed case \cite{DHF13a}. 
 
One can postulate a correlation between the model and physiological phenomena as follows:
 Maintaining cardiac output (CO) is essential for normal function.
 The CO is a product of heart rate and stroke volume.
 In DHF patients, the stroke volume remains unchanged  \cite{Zile,Kitzman}
and hence slowing heart rate should be prevented, or CO will be reduced.
 However, when our model is considered,
 if LV volumes are to be maintained despite increased stiffening (again, as is the clinical finding), LV pressures have to be increased or heart rate decreased. It is thus concluded that the increased LV pressures
 are the result of a forced compensation,
 aimed at maintaining heart rate and CO when the LV wall is stiffened. 

Finally, we note that while the model
 presented here is very much the minimal model for LV function,
 additional phenomenological details, such as ventricular pressure- volume loops
 and atrial parameters may be added in future studies. Such additions
 would render the model analytically unsolvable, yet numerical solutions may be possible,
 and may advance our understanding of different physiological phenomena.

%\bibliographystyle{plain}
%\bibliography{all}

\end{document}